\documentstyle[pra,aps]{revtex}

\input epsf

\newcommand{\qb}{\begin{equation}}
\newcommand{\qe}{\end{equation}}

\begin{document}

\twocolumn[
\title{Diffusion in normal and critical transient chaos}
\author{Z. Kaufmann$^{1}$, H. Lustfeld$^{2}$, A. N\'emeth$^{1,3}$ and 
P. Sz\'epfalusy$^{1,4}$}
\address{
 $^1$ Institute for Solid State Physics,
E\"otv\"os University,
M\'uzeum krt. 6-8, H-1088 Budapest, Hungary\\
 $^2$ Institut f\"ur Festk\"orperforschung,
Forschungszentrum J\"ulich,
D $52425$ J\"ulich, Germany\\
 $^3$ Brody Research Center, GE Lighting Tungsram Co.\ Ltd.,
V\'aci \'ut 77, H-1340 Budapest, Hungary\\
 $^4$Research Institute for Solid State Physics,
P.O. Box 49, H-1525 Budapest, Hungary\\
}
\maketitle

\mediumtext
\begin{abstract}
In this paper we investigate deterministic diffusion in systems which
are spatially extended in certain directions but are restricted in size 
and open in other directions, consequently particles can escape.
We introduce besides the diffusion coefficient $D$ on the
chaotic repeller a coefficient ${\hat D}$ which measures the broadening
of the distribution of trajectories during the transient chaotic motion.
Both coefficients are 
explicitly computed for one-dimensional models, and they are found to be
different in most cases. We show
furthermore that a jump develops in both of the coefficients 
for most of the initial distributions
when we approach the critical borderline
where the escape rate equals the Liapunov exponent of a periodic orbit.
\end{abstract}
] 

\narrowtext
\section{Introduction}

Diffusion is one of the most common and most important 
phenomena of nature. In physics
it is important for nondeterministic 
systems, governed by noise, for quantum systems
and for deterministic systems, as well
\cite{Ha83,AsMe76,Ar94,GeNi82,ScFrKa82,GrFu82,GaNi90,CvEcGa95,TeVoBr96}.
In this paper we want to investigate  transient diffusion in
deterministic systems. By this we
mean diffusion in a system spatially extended in certain 
directions
which is however not closed in other directions
and particles can escape.
The diffusion is usually defined in the limit of infinitely long times.
However, the intrinsic feature of transient 
chaos (see for a review \cite{Te90,CsGySz93})
is the limited time a system is 
close to a chaotic repeller. Only during this time the system shows the
characteristics we are interested in. One can have in mind, for instance,
a  channel in a two-dimensional
mesoscopic system containing Lorentz-type scatterers
which is infinite in one direction but has a finite
width in the transverse direction. Sooner or later all 
particles
are scattered away from the channel and
get lost if we assume free boundaries. 
During their motion in the channel the particles are scattered
and can exhibit diffusion in the extended direction. 
So the question arises in which way
this diffusion should be characterized.

The basic problem is how to define the diffusion coefficient.
One possibility is to compute the diffusion coefficient $D$ for 
particles
on the repeller.  That requires 
the computation of the diffusion coefficient
for trajectories of those particles staying for ever in the channel.
Following this line we can go to
the limit of infinite time and thus can proceed as
in the permanent chaos case.
This formulation, however, can not provide a full description, since the 
probability is zero that a trajectory is on the repeller.
A quite different view is to compute the actual properties 
in the neighborhood of the repeller, which is the main goal of the
present paper.

\begin{figure}
\vspace{-2mm}
\epsfbox{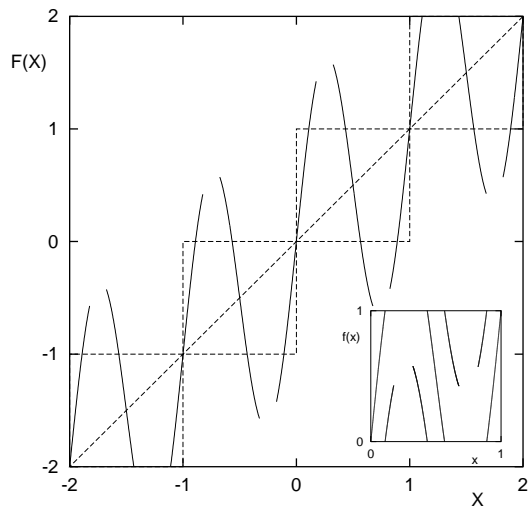}
\caption{An example for map $F$ with translational symmetry and
escape.
The inset shows the corresponding reduced map $f$.}
\end{figure}
To avoid complications with continuous time and higher dimensions
we discuss here one-dimensional discrete 
maps defined on the real line.
The extension of our formulation to more general cases is 
straightforward.
The map $X_{t+1}=F(X_t)$ has translational symmetry
$F(x+1) =F(x) +1$
and is not defined for a periodic set of intervals, called in the 
following windows, from which the trajectory escapes (Fig.\ 1).
It is assumed that the map is chaotic in the sense that
there exists no stable periodic orbit.
Such maps
can be described by reduced maps defined in 
${\cal U}=[0,1]$ 
\cite{GeNi82,ScFrKa82,GrFu82}.
Let us denote by $N$ the box of the interval $[N,N+1)$.
A phase point (called particle in the following) being after
$t$ iterations in box $N$ with position $X_t$
can either remain in that box or move to one of the nearby
boxes or get lost.
Which of these possibilities occurs
depends only on fractional part $\{X\}$ of $X$, since using the new 
variables $N=[X], x=\{X\}$ we get a dynamical system of the two 
variables
\qb
x_{t+1}=f(x_t),\;\;
N_{t+1}=N_t+\Delta(x_t),
\qe
which is equivalent to the map $F$. Here $f(x)=\{F(x)\}$ is the reduced 
map (cf.\ Fig.\ 1), which has a window where trajectories escape, and 
the integer part $\Delta(x)=[F(x)]$ determines the steps among the 
boxes.
We call a system with escape open, otherwise closed. The reduced maps
with escape are generalizations of those introduced in 
\cite{GeNi82,ScFrKa82,GrFu82} and have been studied also in a different
context \cite{Ma95}.

Let us follow the reduced trajectories of particles
in the interval ${\cal U}$. At the beginning the distribution of the 
initial
points $x_0$ is according to an arbitrary density $P_0$ in
${\cal U}$.
The change of $P_t$ after each step of iteration
is induced by $f$ alone and
given by the Frobenius-Perron operator
$P_{t+1}(x)={\cal L}P_t(x)\equiv \sum_{z:f(z)=x}P_t(z)/|f'(z)|$.
Repeated application of ${\cal L}$
leads to an average decrease of the density and to a change of its
structure at the same time. Finally a  quasistationary state is
reached: the density
 normalized in ${\cal U}$ after each step has converged
to $P$, the density of the
 conditionally invariant measure 
\cite{PiYo79},
whereas the density
itself is decreasing after each iteration
by a certain rate, related to the escape rate $\kappa $:
\qb
{\cal L}P =\lambda_cP,\;\; |\ln\lambda_c| =\kappa .
\qe
This emphasizes the importance of the
quasistationary state. Computing the properties of that state 
corresponds to taking the  
limit of large times $T$ to reach it.
We can define a coefficient ${\hat D}$ in the 
quasistationary state, which is analogous to a usual diffusion
coefficient in the sense that it measures the mean quadratic
deviation of the trajectories from their average shift.
We call ${\hat D}$ the {\em transient 
diffusion coefficient}. In  general ${\hat D}$ differs from
$D$ as will be shown explicitly for 1D models.
 
We study the properties of both diffusion coefficients as a 
function of a control parameter while keeping the escape rate constant.
Within the family of such maps there is a limit when the Liapunov 
exponent of one of the unstable periodic orbits has reached the value 
of the escape rate and still the conditionally invariant measure has a 
density \cite{NeSz95}.
This situation is called  the 
critical borderline of transient chaos \cite{NeSz95,LuSz96}.
We show that here
both $D$ and ${\hat D}$ can have two different values depending
on the initial 
distribution of trajectories. This will be shown to be traced back to
the existence of two conditionally invariant densities with different
escape rates.

In Section 2 we derive the diffusion coefficient on a
repeller, $D$,  and 
the transient diffusion coefficient ${\hat D}$.
Then we choose maps $F$ of open systems whose reduced maps
are in the family of fully developed chaotic 
maps\cite{GySz79,NeSz95}. We
present results that
clearly demonstrate the difference between $D$ and ${\hat D}$.
In Section 3 we show that on approaching the critical borderline of 
transient chaos
a jump develops in both $D$ and ${\hat D}$ for most of the
initial distributions. The conclusion ends the paper.

\section{The diffusion coefficients of transient chaos}

To get the diffusion coefficient for open systems we have to
know the shift $S_t$ of particles among the boxes at a given time 
\qb
S_t= N_t-N_0= \sum_{k=0}^{t-1} \Delta(x_k).
\qe
Let us compute the average value of $S_t$ first
by averaging over an initial distribution $P_0(x_0)$ of $x_0$ 
values with $f^{t}(x_0)$ in ${\cal U}$.
Thus we obtain
$
{\langle S_t \rangle}_0 =
{\sum_{l =0}^{t-1}\int_{{\cal U}_{t}} \Delta (f^l(x))P_0(x) dx}/
{\int_{{\cal U}_{t}}P_0(x)dx}
$
with
$
{\cal U}_t =f^{-t}({\cal U}) \cap {\cal U}
$.
For $t \rightarrow \infty\;$
${\cal U}_t$ becomes a repelling invariant set of the reduced map
which will be called repeller in the following.
To get the diffusion coefficient the procedure is to
take the limit $t \rightarrow \infty $, which restricts
the trajectories on the repeller.
Apart from that the 
derivation of the corresponding expressions becomes 
straightforward along the lines of permanent diffusion
\cite{ScFrKa82,GrFu82}.
We use the Frobenius-Perron operator ${\cal L}$ to write
$
{\langle S_t \rangle}_0 ={\sum_{l =0}^{t-1}\int_{\cal U}
[{\cal L}^{t-l}\Delta {\cal L}^l P_0](x)dx }/
{\int_{\cal U}[{\cal L}^{t} P_0](x)dx}
$.
But $\lim_{l \rightarrow  \infty } \lambda_c^{-l}{\cal L}^l$ 
acts as a projector on $P$, the 
density of the conditionally invariant measure.
Thus the result for the averaged drift is
\qb
\lim_{t\rightarrow\infty} \frac{1}{t}{\langle S_t\rangle}_0
=\int_{\cal U} \Delta(x)\,\mu(dx)
\equiv{\langle\Delta\rangle}_\mu,
\qe
where $\int_{\cal U}\,\mu(dx)=1$ and
$\mu$ is the natural invariant measure on the repeller.
${\langle S_t^2\rangle}_0$ can be computed in an analogous manner.
The diffusion coefficient is then defined by
\qb
D =\lim_{t\rightarrow\infty}
\frac{{\left\langle(S_t-{\langle S_t\rangle}_0)^2\right\rangle}_0}{2t},
\qe
which can be written in terms of
the correlation function $C_{\Delta\Delta}$ 
 on the repeller as
\qb
D=\lim_{t\rightarrow\infty}\frac{1}{2t}
\sum_{l,m=0}^{t-1} C_{\Delta\Delta}(|l-m|),
\qe 
where
$C_{\Delta\Delta}(l)=
{\left\langle(\Delta(f^l(x))-{\langle\Delta\rangle}_\mu)
\Delta(x)\right\rangle}_\mu$.
The expressions are  
completely analogous to the formulae 
of closed systems \cite{ScFrKa82,GrFu82}.
In spite of the analogies
${\langle\Delta\rangle}_\mu$ and $D$
cannot be accepted as the
only quantities characterizing open systems
since a trajectory moves on the repeller with probability zero.

Because of this a second procedure is suggested.
Let us investigate particles that have been injected into the system 
and
have survived (i.e.\ have not escaped) after a time $T$ long enough to
have the density $P_T(x)$ of the reduced distribution  of these 
particles close to the density $P(x)$ of
the conditionally invariant measure. 
We consider the shift of the particles from time $T$ until their escape
$S_e(x_T)=N_{T+\tau(x_T)}-N_T=\sum_{k=0}^{\tau(x_T)-1}
\Delta(x_{T+k})$,
where
$\tau(x)$ is the number of steps taken by the trajectory of $x$,
i.e.\ $f^\tau(x)\in{\cal U}\equiv [0,1]$ and
$f^{\tau(x)+1}(x)\not\in{\cal U}$.
The average of this shift
${\langle S_e\rangle}_T=\int_{\cal U} S_e(x)P_T(x)\,dx$
over the trajectories considered and the average lifetime 
${\langle\tau\rangle}_T=\int_{\cal U} \tau(x)P_T(x)\,dx$
can be evaluated in the limit $T\rightarrow\infty$
using properties of the Frobenius-Perron operator.
The result is $\langle\tau\rangle={\lambda_c}/{(1 -\lambda_c)}$
and an average drift
\qb
\frac{\langle S_e\rangle}{\langle\tau\rangle}=\langle\Delta\rangle\equiv
\frac{1}{\lambda_c}\int_{{\cal U}_1}\Delta(x)P(x)\,dx.\label{avdt}
\qe
For sake of simplicity the $\infty$ index at the sign of average
is omitted.

We can characterize the transient spreading of the particles
by the transient diffusion coefficient defined as
\qb
\hat D=\lim_{T\rightarrow\infty}
\frac{{\langle\hat S_e^2\rangle}_T}{2{\langle\tau\rangle}_T}=
\frac{\langle\hat S_e^2\rangle}{2\langle\tau\rangle},
\qe
where
$\hat S_e(x_T)=\sum_{k=0}^{\tau(x_T)-1}
(\Delta(x_{T+k})-\langle\Delta\rangle)$,
the corrected shift is introduced. Through steps similar to those
which led to Eq.\ (\ref{avdt})
$\hat D$ can be related to
the correlation function of transient chaos\cite{LuSz96}
$c_{\Delta \Delta }(n) =\lambda_c^{-(n+1)}\int_{{\cal U}_{n+1}}
(\Delta(f^n(x))-\langle\Delta\rangle) \Delta(x)P(x)\,dx$
by the expression
\qb
{\hat D} = \sum_{n=-\infty}^\infty \lambda_c^{|n|}
c_{\Delta\Delta}(|n|).
\qe
Physically this means the following. After injection drift and spreading
of the injected particles depend on the initial distribution, but its 
influence
decreases exponentially fast. After a long time $T$ it is no longer
important and then further drift and spreading of the particles is 
dictated by the properties of the open system alone, characterized by
$\langle S_e\rangle$ and ${\hat D}$, respectively. This general picture
has to be refined, however, if the system has more 
than one smooth conditionally
invariant measure as will be seen later.

In the numerical calculations of open systems we take the reduced map
$f$ from the family of maps defined on the interval $[0,1]$
and given by the inverse branches
$f_l^{-1}(x) ={(x+v(x))}/{2R}$ (lower branch),
$f_u^{-1}(x) =1-f_l^{-1}(x)$ (upper branch),
where $R\ge 1$, $v(0)=0$, $v(x)$ is symmetric and $|v'(x)|\le 1$.
The density of a conditionally invariant measure can be expressed as
\qb
P(x) = 1 +v'(x). \label{invm}
\qe
The width of the window and the escape rate are given by $1-R^{-1}$ and
$\kappa=\ln R$, respectively.
Note that in the limit $v\equiv 0$ the open tent  map emerges whereas in
the limit $v'(0)=1$ we reach the state what we call critical
borderline of transient chaos, since the Liapunov exponent of the
fixed point at the origin equals the escape rate \cite{NeSz95,LuSz96}.

In our numerical computations we take ${\cal U}=[0,1]$ and
we use the piecewise parabolic map\cite{NeSz95,GySz79} 
which corresponds to
$v(x) =d\cdot x(1-x)$, $-1\le d\le 1$.
We require, that a particle hitting the left (right) branch of $f$ 
hops to the left (right) adjacent box $N$ in the map $F$.
If it hits neither branch then it escapes.
This yields
$\Delta (x)=-1$ if $x<{1}/{(2R)}$ and $1$ if $x\ge 1-{1}/{(2R)}$.
To be distinctively in the repeller regime we 
choose $R=1.5$. However, the 
computation of $D$ is not trivial in that regime. 
Reliable results were obtained only
after we had applied backward iteration up to a depth of $20$ steps, 
left out the corresponding
gaps of the repeller and had started with $10^8$ trajectories
($10^9$ in case of $d=0.95$).
While computing ${\hat D}$ we did not take into 
account the first 20 steps ($T=20$) to ensure the
quasistationarity of the system.
In Fig.\ 2 the diffusion coefficients $D$ and ${\hat D}$ are
shown for $d$ values ranging from $0$ to $1$. Whereas they agree for the
tent map a difference between $D$ and ${\hat D}$ develops when the 
parameter $d$ increases. The two extra points for $d=1$ in Fig.\ 2 will
be explained in the next section.
\begin{figure}
\vspace{-2mm}
\epsfbox{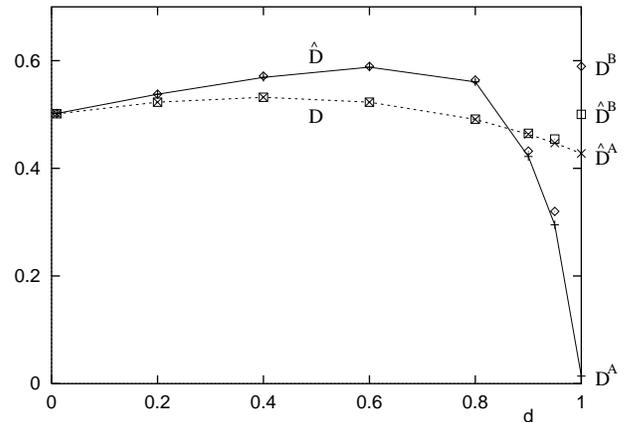}
\caption{Numerical results for $D$ and ${\hat D}$ in function of $d$.
The superscript refers to the choice of the interval on which
the initial points were distributed.}
\end{figure}

\section{Diffusion in critical transient chaos}

A peculiar situation emerges when we reach the critical borderline of
transient chaos ($d=1$ in the example treated). Eq.\ (\ref{invm}) is
still valid and gives a density $P^A(x)$ (equal to $2(1-x)$ 
in case of our example) for a conditional
invariant measure with an escape rate $\kappa^A=\ln R$. The reason
for introducing the superscript $A$ is explained below. The
corresponding eigenfunction $P^{+A}$
of the adjoint Frobenius-Perron operator consists of two
$\delta$-function peaks $P^{+A}=
{(\delta(x)+\delta(1-x))}/2$ \cite{LuSz96}. This makes possible 
the existence of an
other conditionally invariant measure with density. To find it one should
iterate a function $\phi_0$ that is positive everywhere in $(0,1)$ and
$\phi_0(0)=\phi_0(1)=0$ by the Frobenius-Perron operator with the prescription
$\phi_{t+1}=\alpha_t{\cal L}\phi_t$, where $\alpha_t$ is chosen in such a
way that the norm of $\phi_{t+1}$ is kept to unity. The iterated
functions remain orthogonal to $P^{+A}$ because in the critical state
the derivative of the map at the maximum points is
infinity. Furthermore it is easy to see that the positivity of
$\phi_t$ is also maintained.
In the limit $t\rightarrow\infty$ the iteration leads to a nonnegative
eigenfunction which can be taken as the density $P^B(x)$ of a
conditionally invariant measure. The obtained eigenvalue $\lambda_c^B$
defines a new escape rate $\kappa_B=|\ln\lambda_c^B|$.

This means that we have two
different conditionally invariant measures with different escape rates
in the critical state. Which of them is obtained in the long time
limit when iterating the Frobenius-Perron operator depends on the
initial density $P_0(x)$. It is clear from the above considerations
that starting with a smooth $P_0(x)$ such that $P_0(0)=P_0(1)=0$ we
reach $P^B(x)$, otherwise $P^A(x)$. When simulating the initial
density by a large number of initial points the obtained
distribution $P_T(x)$ of iterates at a large time $T$ is governed by
the same rule. Note that to balance the escape we have to increase the
number of initial points when increasing $T$.

The above rule for the asymptotic distribution is reflected in the
values of the diffusion coefficients $D$ and ${\hat D}$. In the
numerical calculations we distributed a large number of particles
uniformly on intervals, $[0,1]$ and $[0.1,0.2]$, to simulate initial
densities in the basin of attraction of $P^A$ and $P^B$,
respectively. It is clear from Fig.\ 2 (at $d=1$) that the result
depends strongly on the initial distribution
(as well as values of the escape rate and the average drifts
${\langle\Delta\rangle}_\mu$, ${\langle\Delta\rangle}$, not displayed here).
Note, the value of $D^A$
is zero due to the fact that the natural measure induced by $P^A$ on
the repeller (but not the repeller itself) degenerates,
it becomes a Dirac
delta function at the origin\cite{NeSz95,LuSz96}.

\section{Discussion}

In this paper we have investigated statistical properties in long time
limit of trajectories in the transiently chaotic regime by the help of
two diffusion coefficients $D$ and ${\hat D}$.
For $D$ this time limit
refers to the whole length of the trajectories (a procedure leading 
automatically to experience the properties of the repeller itself).
In contrast the long time limit of  
${\hat D}$ is only needed to measure it in the quasistationary state.
Though the definition of a transient diffusion constant as introduced
in the present paper seems to be the most natural one, it is not
the only possibility. In future work we plan to examine also alternative
definitions. 

We have shown furthermore that a map at the critical borderline of 
transient chaos has two smooth conditionally invariant measures. 
We find, however, that in nearly all practical situations
(i.e.\ when the initial phase points are concentrated in the inner
part of the interval)
the relevant  measure is not the usual one, belonging to the smaller
escape rate, but an other
eigenfunction of the Frobenius-Perron operator ${\cal L}$.

In this paper we have treated a 1D model which can be conceived as 
describing
the dynamics along the expanding direction of a 2D map. It is
expected that a more complete treatment would not change the
conclusions.
 We plan as a next step to study 2D
Hamiltonian systems.

\acknowledgments
This work has been supported in part by the 
German-Hungarian Scientific and Technological Cooperation 
{\em  Investigation of classical and quantum chaos},
by the Hungarian National Scientific Research Foundation under Grant 
Nos.\ OTKA T017493 and OTKA F17166,
the US-Hungarian Science and Technology Joint Fund in cooperation
with the NSF and the Hungarian Academy of Sciences under project
No.\ 286,
and the Ministry of Education of Hungary under grant No.\ MKM337.
One of the authors (H.\ L.) would like to thank Prof.\ P. Sz\'epfalusy
for the hospitality at E\"otv\"os University where this work
has been done.

\end{document}